\documentclass[preprint]{imsart}

\RequirePackage[OT1]{fontenc}
\RequirePackage{amsthm,amsmath}
\RequirePackage{natbib}
\RequirePackage[colorlinks,citecolor=blue,urlcolor=blue]{hyperref}

\usepackage{graphicx}
\usepackage{amssymb}
\usepackage{url}
 
\renewcommand{\Pr}{\mathsf{P}}

\newcommand{\given}{\mid}

\def\BF{\rm BF}

\startlocaldefs
\numberwithin{equation}{section}
\theoremstyle{plain}

\endlocaldefs

\begin{document}

\begin{frontmatter}
\title{Wavelet-based genetic association analysis of functional phenotypes arising from high-throughput sequencing assays}
\runtitle{Wavelet-based association analysis of functional data}

\begin{aug}
\author{\fnms{Heejung} \snm{Shim}\ead[label=e1]{hjshim@gmail.com}}
\and
\author{\fnms{Matthew} \snm{Stephens}\ead[label=e2]{mstephens@uchicago.edu}}

\runauthor{Shim and Stephens}

\affiliation{University of Chicago}

\address{Heejung Shim\\
Departments of Human Genetics\\
University of Chicago\\
920 E. 58th Street, CLSC 5th floor\\
Chicago, IL, 60637, U.S.A.\\
\printead{e1}\\
}

\address{Matthew Stephens\\
Departments of Statistics and Human Genetics\\
University of Chicago\\
920 E. 58th Street, CLSC 5th floor\\
Chicago, IL, 60637, U.S.A.\\
\printead{e2}\\
}


\end{aug}

\begin{abstract}
Understanding how genetic variants influence cellular-level
processes is an important step towards understanding how they influence important organismal-level traits, or ``phenotypes", including human disease susceptibility.
To this end scientists are undertaking large-scale genetic association studies that aim to identify genetic variants associated with molecular and cellular phenotypes, 
such as gene expression, transcription factor binding, or chromatin accessibility. 
These studies use high-throughput sequencing assays (e.g.~RNA-seq, ChIP-seq, DNase-seq)
to obtain high-resolution data on how the traits vary along the genome in each sample.
However, typical association analyses 
fail to exploit these high-resolution measurements, instead aggregating the data at coarser resolutions, such as genes, or windows of fixed length. 
Here we develop and apply statistical methods that better exploit the high-resolution data.
The key idea is to treat the sequence data as measuring an underlying ``function" that varies along the genome,
and then, building on wavelet-based methods for functional data analysis, test for association between genetic variants and the underlying function.
Applying these methods to identify genetic variants associated with chromatin accessibility (dsQTLs) we 
find that they identify substantially more associations than a simpler window-based analysis, and in total we identify 
772 novel dsQTLs not identified by the original analysis. 
\end{abstract}



\end{frontmatter}

\section{Introduction}
Genetic association studies aim to understand the function of genetic
variants by associating them with observable traits, or ``phenotypes". 
Although many association studies have focused on organismal-level phenotypes, such as human disease (e.g. \citet{WTCCC_2007}), association studies also provide a powerful tool for studying molecular-level phenotypes, such 
as gene expression \citep{Cheung_2010, Pickrell_2010,Montgomery_2010}, transcription factor binding \citep{Kasowski2010, Karczewski2013} and chromatin accessibility \citep{Degner_2012}.
Measurement of many molecular phenotypes has been recently transformed by the advent of cheap high-throughput sequencing technology, and corresponding experimental protocols (RNA-seq \citep{Mortazavi_2008a,Wang_2008,Marioni_2008}, ChIP-seq \citep{Johnson_2007, Barski_2007, Mikkelsen_2007}, DNase-seq \citep{Boyle_2008, Hesselberth_2009}), which provide high-resolution measurements across the whole genome. However, typical analyses fail to exploit these high-resolution measurements, instead aggregating the data at coarser resolutions, such as genes, or windows of fixed length. 

In this paper we develop and apply association analysis methods 
that better exploit high-resolution measurements from high-throughput sequencing assays. We specifically focus on identifying
genetic variants that are associated with an epigenetic phenomenon known as chromatin accessibility, measured using DNase-seq \citep{Boyle_2008, Degner_2012}, both described in more detail below. However, the same or similar ideas could also be applied to association analyses of other high-throughput sequencing measurements. 

Conceptually, the key idea is to treat the data from high-throughput sequencing assays as noisy measurements of an underlying ``function" (in this case, chromatin accessibility) that varies along the genome. We then adapt methods from functional data analysis, based on wavelets, to develop a test for association between a covariate of interest (in this case, a genotype) and the shape of the underlying function. We also
provide methods to estimate the shape of the genotype effect, which can help in understanding the potential mechanisms underlying the identified associations.

In outline, our methods first transform the
data using a wavelet transform, and then  
model associations in the transformed space, rather than the original data space. This approach makes modeling easier because we expect the
effect of genotype on phenotype to exhibit a spatial structure in the original space, which corresponds to a sparse structure in the transformed space, and sparsity is relatively easy to model. Here we are borrowing ideas that have been developed, more generally, in the ``functional mixed models"
work of \citet{Morris_Carroll_2006, Morris_2008, Zhu_2011}. In particular, \citet{Morris_2008} presented a framework for identifying locations within a region that show significant effects of covariates. Other relevant work on wavelet methods for regression analysis of functional data include  \citet{Fan_Lin_1998, Abramovich_Angelini_2006, Antoniadis_Sapatinas_2007,Zhao_2008, Yang_2008}. Previous applications of wavelet-based methods
in genomics, include \citet{Spencer2006, Day2007, Zhang2008, Wu2010, Mitra2012,Clement2012}. 
Our main contributions are to
embed the wavelet-based methods into a framework for association testing that is computationally
tractable for large-scale genetic association analyses that involve hundreds of thousands of tests, and to
demonstrate the practical potential of these methods for associating genetic variants with sequence-based molecular phenotypes.


\section{Background}

\subsection{DNase-seq and chromatin accessibility}

In brief, DNase-seq is an experimental protocol that measures the accessibility, or openness, of {\it chromatin} along the genome. 
Chromatin consists of both the DNA that makes up the genome and the proteins that package it within the cell nucleus.
Accessibility is important because it is associated with biological function, and DNase-seq has been a useful tool for detecting functional elements of the genome \citep{Boyle_2008}.  Chromatin accessibility at any given location will vary from cell to cell, and although single-cell experiments are on the horizon, almost all current experiments provide average measurements over a population of cells, usually from the same individual.

The key step in the DNase-seq protocol is the use of an enzyme called DNase I to selectively cut the DNA at locations where the chromatin is accessible.  There is a quantitative aspect to this selection: other things being equal, locations where the chromatin is more accessible will tend to be cut more often. The locations of these cut points are revealed by sequencing the ends of the resulting fragments of DNA, and mapping the sequences (the ``reads") back to the genome. The resulting data are then conveniently summarized by the counts, $c_b$, of the number of cut points at each base in the genome (for humans, $b \approx 1,\dots, 3 \times 10^9$). (Note that 
$c_b$ denotes the number of reads that {\it start} at base $b$, rather than the number of reads that {\it cover} base $b$, so each read is counted only once.)
In analyses these counts are usually standardized to account for the total number of sequence reads generated for each sample, so we here use $d_b = c_b / S$
where $S$ is the total number of mapped reads in the experiment.
Although the process is
subject to considerable technical variation, and other confounding factors, higher values of $d_b$ generally correspond to higher accessibility of base $b$. (Technically, the DNase-seq protocol actually measures ``DNase I sensitivity", or sensitivity 
to cutting by the DNase I enzyme, which is a proxy for chromatin accessibility. For simplicity we ignore this distinction here.)

A typical experiment will produce millions of sequence reads per sample, and these will be concentrated in the relatively small proportion of the genome that is most ``accessible". Thus $d_b=0$ for most bases $b$, but some regions will show substantial counts at each base. 
Further, where it exists, accessibility
 tends to extend over hundreds of bases, and more generally $d$ tends to exhibit local spatial autocorrelation (``spatial structure"). 
  One important goal of our methods is to account for this structure in the analysis.
 
Here we consider data from \citet{Degner_2012}, who collected DNase-seq data on samples from 70 different human individuals, for whom extensive genome-wide
genetic data are also available. By correlating the DNase-seq data with the genetic data, we aim to \emph{identify genetic variants
associated with chromatin accessibility}. Such genetic variants are referred to as
dsQTLs (DNase I sensitivity Quantitative Trait Loci) by \citet{Degner_2012}.
Identifying genetic variants that are associated with chromatin accessibility, and other
molecular phenotypes such as transcription factor binding and gene expression, can help provide insights into the mechanisms by which genetic variation influences
gene regulation. Indeed, \citet{Degner_2012} found that many of the dsQTLs they identified were also associated with gene expression (which is associated with
protein production), suggesting that genetic variation affecting transcription factor binding and chromatin accessibility may 
explain a substantial proportion of genetic variation in protein production. Ultimately, by combining these types of data on molecular-level phenotypes,
and integrating them with similar data on organismal level phenotypes,
we hope to understand which genetic variants affect human disease susceptibility, and the biological mechanisms by which they operate \citep{Nicolae2010}.
Identifying dsQTLs, as we do here, is one helpful step towards this larger goal.

\subsection{Wavelets}

Wavelets are a tool from signal processing that are commonly used to deal with spatially-structured (or temporally-structured) signals.
In this paper we use the Haar Discrete Wavelet Transform (DWT), and this section
provides a brief intuitive description of the DWT. Further, more formal, background on wavelets can be found in \citet{Mallat_1989}. 

Let $d =(d_b)_{b=1}^B$ be the standardized counts from a DNase-seq experiment in a region with a length $B$ assumed to be a power of 2 ($B=2^J$). The DWT decomposes $d$ into a series of ``wavelet coefficients" (WCs), $y = (y_{sl})$,
each of which summarizes information in $d$ at a different scale  (or resolution) $s$ and location $l$.
At the ``zeroth scale" there is a single WC ($y_{01}$), 
which is simply the sum of the elements of $d$,  $y_{01} = \sum_b d_b$. (This ``zeroth scale" WC is not truly a WC, but we use this shorthand here for convenience.)  This coefficient summarizes $d$ at the coarsest possible level, by its sum. 
At the first scale there is also a single WC ($y_{11}$), which contrasts the counts in the first half vs second half of the region.  That
is $y_{11}:= \sum_{b \leq B/2} d_b - \sum_{b>B/2} d_b$ (omitting a scaling constant that is usually used to normalize the WCs, but does not concern us here). This WC can be thought of as roughly capturing any trend in $d$ across the region.
At the second scale there are two WCs ($y_{21},y_{22}$): the first contrasting the first quarter vs the second quarter of the region; and the second contrasting the third quarter vs the fourth quarter of the region. This process continues through the scales: at scale $s$ there are $2^{s-1}$ WCs that contrast regions of length $2^{J-s}$, and hence capture higher-resolution features of $d$. 

Since $y$ is a linear transform of $d$, the DWT can be written as a matrix multiplication: $y = Wd$ where
$W$ is known as the DWT matrix. Further, the transform is one-one, so $W$ is invertible, and  $d$ can be obtained from $y$ by 
the ``inverse discrete wavelet transform" (IDWT), $d = W^{-1}y$. We exploit this linearity of the IDWT later to obtain closed form 
expressions for posterior mean and variances of effect sizes in the original scale (see Methods).

Because the WCs are simply a one-one transform of $d$, $y$ contains exactly the same information as $d$.
However, WCs have two crucial properties that make them useful for settings where, as here, $d$ is expected to have a spatial correlation structure: 
i) where values of $d$ may be strongly spatially correlated, the WCs tend to be less dependent, referred to as the ``whitening" property of the wavelet transform; ii) Typically, many WCs will be small, with the signal concentrated in a few ``big" WCs.  As a result one can
obtain denoised (smoothed) estimates of a signal by ignoring or shrinking the smaller WCs (i.e. reducing them towards 0). 
This is called ``wavelet denoising" \citep{Donoho1995}. Here we effectively apply wavelet denoising to estimate
the effect of a genetic variant on a signal, rather than to the signal itself (see also \citet{Morris_Carroll_2006, Zhu_2011} for example).



\section{Methods}

Our data consist of DNase-seq data and genotype data at genetic variants (mostly Single Nucleotide Polymorphisms, or SNPs) across the whole genome on $N$ individuals,
and our goal is to assess whether the DNase-seq data is associated with the genotype data.
In practice we expect that SNPs affecting chromatin accessibility will tend to have a relatively local effect,
an expectation supported by results in \citet{Degner_2012}. Thus, similar to \citet{Degner_2012}, we first divide the DNase-seq data
into regions (of length $B=1024$ in this case; see Results), and then test each region for association with all near-by SNPs. 
We will first describe the test for a single SNP, and then describe how we apply it to test all near-by SNPs.

Let $d^i$ denote the vector of DNase-seq count data for individual $i$ ($i=1,\dots,N$). Thus $d^i$ is a vector of counts of length $B=2^J$.
Let $g^i$ denote the genotype data for individual $i$ at a single SNP of interest,
 coded as 0, 1, or 2 copies of the minor allele (so $g^i \in \{0,1,2\}$).
Our aim is to assess whether the DNase-seq data is associated with genotype at this SNP.
That is, 
can we reject the null hypothesis $H_0$ that $d$ is independent of $g$?

In outline, our approach is as follows. First, we transform each phenotype vector $d^i$ using the DWT outlined above, to produce
a new phenotype vector $y^i$ of wavelet coefficients (WCs). 
Then, based on simplifying modeling assumptions detailed below, 
which combine information across WCs into a hierarchical model, we compute a likelihood-ratio test statistic $\hat\Lambda$ testing $H_0$. Finally, since the modeling assumptions are unlikely to hold exactly in practice, we use permutation to assess significance of the observed value of $\hat\Lambda$.



In more detail, let $y_{sl}$ denotes the vector of WCs at scale $s$ and location $l$, and let $\gamma_{sl}$ denote a binary indicator for whether $y_{sl}$ is associated with $g$.
The null hypothesis, $H_0$, is that there is no association between any WC and $g$; that is, $\gamma_{sl}=0$ for all $s$ and $l$.

To measure the support for $\gamma_{sl}=1$ for a specific $s,l$ we use a Bayes Factor,
\begin{equation}
\BF_{sl}(y, g) := \frac{p(y_{sl} | g, \gamma_{sl}=1)}{p(y_{sl} | g, \gamma_{sl}=0)}.
\end{equation}
To compute this Bayes Factor we use the models and priors from \citet{Servin2007},
which are based on assuming a standard normal linear regression for $p(y_{sl} | g, \gamma_{sl})$: 
\begin{equation} \label{y:eqn}
y_{sl}^i = \mu_{sl} + \gamma_{sl} \beta_{sl} g^i + \epsilon_{sl}^i   \quad \text{with}  \quad \epsilon_{sl}^i \sim \mathcal{N}(0,  \sigma_{sl}^2),
\end{equation}
where $\mu_{sl}$ denotes the mean WC of individuals with $g^i=0$; $\beta_{sl}$ denotes the effect size of $g$ on the WC; and $\epsilon_{sl}^i$ is the residual error for sample $i$. With appropriate priors on $\mu_{sl}, \beta_{sl}, \sigma_{sl}$ (see Appendix~\ref{DetailsBF}) the Bayes Factor $\BF_{sl}$ has a simple analytic form. To reduce the influence of deviations from the normality assumption, we quantile transform the vector of WCs, $y_{sl}$, to the quantiles
of a standard normal distribution, and compute $\BF_{sl}$ using the transformed WCs.

To combine information across scales $s$ and locations $l$ we build a hierarchical model for the $\gamma_{sl}$,
assuming
\begin{equation}
p(\gamma_{sl}=1 | \pi) = \pi_s
\end{equation}
where $\pi=(\pi_1,\dots,\pi_J)$ is a vector of hyperparameters, with $\pi_s$ representing the proportion
of WCs at scale $s$ that are associated with $g$. 
Then, assuming independence across scales and locations, the likelihood ratio for $\pi$, relative to $\pi \equiv 0$ (i.e.~$\pi_s = 0$ $\forall s$), 
is given by
\begin{align}
\Lambda(\pi; y,g) := \frac{p(y | g, \pi)}{p(y| g, \pi \equiv 0)} & = \prod_{s,l} \frac{p(y_{sl} | g, \pi_s)}{p(y_{sl} | g, \pi_{s}=0)} \\
&= \prod_{s,l} \frac{\pi_s p(y_{sl} | g,\gamma_{sl}=1) + (1-\pi_s) p(y_{sl} | g, \gamma_{sl}=0 )}{p(y_{sl} | g, \gamma_{sl}=0)} \\
&= \prod_{s,l} [\pi_s \BF_{sl} + (1-\pi_s)]. \label{eqn:LR0}
\end{align}

Within this hierarchical model, the null $H_0$ holds if $\pi \equiv 0$. Thus, to test $H_0$ 
we use the likelihood ratio test statistic 
\begin{equation}
\hat\Lambda(y,g):= \Lambda(\hat\pi; y,g)
\end{equation}
where $\hat\pi$ denotes the maximum likelihood estimate $\hat\pi :=\arg\max\Lambda(\pi; y, g)$.
This is easily computed using an EM algorithm.
 
Our hierarchical model assumes conditional independence of $y_{s,l}$ (and $\beta_{s,l}$) given $\pi$ across scales and locations.  This assumption is partly justified by the whitening property of the DWT mentioned above; and certainly a corresponding conditional independence assumption would be entirely inappropriate for the original data $d_b$ due to spatial correlations. Nonetheless, the conditional independence assumption will not hold exactly in practice. Anticipating this concern, we note that a primary goal of the hierarchical model is to obtain a test statistic for $H_0$, whose 
 significance is assessed by permutation (see below), and that the resulting $p$ values are valid regardless of the correctness of the modeling assumptions.

\subsection{Multiple SNPs and permutation procedure} \label{permutation} 

The statistic $\hat\Lambda(y,g)$ tests for association between $y$ (or, equivalently, $d$) and a single SNP with genotype vector $g$. Often one would
like to ask, for a given region, whether $y$ ($d$) is associated with {\it any} of many nearby SNPs. 
To assess this for a set of $P$ nearby SNPs, with genotype vectors given by $g_1,\dots,g_P$, we use the test statistic
\begin{equation}
\hat\Lambda_\text{max} := \max_p \hat\Lambda(y,g_p).
\end{equation}

To assess significance of $\hat\Lambda_\text{max}$ we use permutation.
That is, we generate independent random permutations $\nu_1,\dots,\nu_M$ of $(1,\dots,N)$, and compute 
\begin{equation}
\hat\Lambda_\text{max}^j := \max_p \hat\Lambda(y, \nu_j(g_p)).
\end{equation}
Then the $p$ value associated with $\hat\Lambda_\text{max}$ is 
\begin{equation}
p = \frac{\#\{j: \hat\Lambda_\text{max}^j \geq \hat\Lambda_\text{max} \} + 1}{M+1}.
\end{equation}


\subsection{Filtering of low count WCs}

Some WCs, particularly those corresponding to high resolutions, are computed based on very low counts.
Indeed, for some WCs, the majority of individuals have zero counts in the regions being contrasted, and so have a WC of zero.
These WCs effectively have high sampling error, and provide little information on association; however our model (\ref{y:eqn})
does not incorporate the sampling error, and so these WCs tend to contribute more than they should to $\Lambda$, effectively 
adding noise to the test, and reducing power. To address this, we filter out these ``low count" WCs, by setting their $\BF_{sl}=1$
in equation (\ref{eqn:LR0}) (a BF of 1 corresponds to no information about association). In results presented here, a set of WCs $\{y^i_{sl}\}_{i=1}^N$
was considered ``low count" if the total counts used in their computation were less than two per individual on average (i.e. $<140$ in our data with 70 individuals).

\subsection{Controlling for confounding factors} \label{normalization}

In genetic association analyses of molecular-level phenotypes, power can be substantially increased by controlling for unmeasured confounding factors
\citep{Leek2007, Stegle2010}. In this setting, this can be achieved by estimating the unmeasured factors by Principal Components Analysis,
and then regressing out the first few Principal Components (PCs) from the phenotypes before testing them for association with genotype. (In our data analysis
here we use the four PCs used by \cite{Degner_2012}.)
Specifically our procedure is as follows. After quantile transforming each WC to a standard normal distribution, we correct these transformed WCs by taking the residuals
of a standard multiple linear regression of the WCs on the PCs. Finally, we quantile transform these residuals to the quantiles of a standard normal distribution
and use these quantile-transformed residuals in the Bayes Factor calculations. Further data normalization could also be helpful
(e.g.~GC content correction \citep{Pickrell_2010, Benjamini2012}), but we do not pursue
this here.

\subsection{Effect size estimates}

Under the above hierarchical model, given $\hat\pi$, the posterior distributions on the effect sizes in the wavelet space, $p(\beta_{sl} | y,g,\hat\pi)$, are available in closed form.
Specifically, the $\beta_{sl}$ are {\it a priori} independent, each having a distribution that is
a mixture of a point mass at zero and a three parameter version of a $t$ distribution \citep{Jackman2009},
with density given in Appendix~\ref{DetailsEffectSize}. 

However, the effects $\beta_{sl}$ in the wavelet space are not easy to interpret.
To obtain interpretable estimates of the effect of a SNP $g$ we transform these effects from the wavelet space back to the data space using the IDWT.
To explain, we combine the $B$ equations of the form (\ref{y:eqn}) (corresponding to the $B$  values of $s,l$) into a single matrix equation:
\begin{equation} \label{eqn:matrixy}
Y = M + \beta g + E
\end{equation}
where $Y,M$ and $E$ are $B \times n$ matrices (the  WCs, means and residuals respectively), $\beta$ is a $B \times 1$ matrix of effects, and $g$ is a $1 \times n$ matrix of genotypes. Now recall that $D= W^{-1}Y$ where  $W$ is the DWT matrix, so premultiplying (\ref{eqn:matrixy}) by $W^{-1}$ yields
 \begin{equation}
 D = \tilde M + \alpha g + \tilde E
 \end{equation}
where $\tilde M=W^{-1}M$, $\tilde E = W^{-1}E$ and $\alpha:= W^{-1} \beta$ is a $B$ vector of effect sizes in the original data space. 

Thus the effects in the original space, $\alpha$, are given by the IDWT of $\beta$, which is a linear function of $\beta$.
Although the full posterior on $\alpha$ does not have a simple analytic form, the linear relationship
with $\beta$ yields closed forms for the pointwise posterior mean and variance of $\alpha_b$ for $b=1,\dots,B$ (see Appendix~\ref{DetailsEffectSize}).
Here we use these posterior summaries to summarize the posterior distribution on the effects. Other
types of posterior inference could be performed by simulating from the posterior for $\alpha$ (which is easily achieved
by simulating from the posterior of $\beta$ and applying the IDWT to the simulated samples).



\section{Results}

\subsection{The data and previous analysis}
We apply our approach to DNase-seq data from \citet{Degner_2012}, who also used these data to identify dsQTLs. 
We begin with a brief summary of the analysis in \citet{Degner_2012}. The authors collected DNase-seq data for 70 HapMap Yoruba LCLs, and correlated these DNase-seq data with a total of about
18.8 million genetic variants (either directly genotyped or imputed). 
To do this they first identified regions of the genome that had many DNase-seq reads mapping to them, since these are most
likely to contain functional regulatory elements, and are most amenable to association analysis.
(Regions with no reads are clearly not amenable to association analysis.) 
 Specifically, they divided the whole genome into non-overlapping 100bp windows, and took the top 5\% of these windows
 ranked according to a DNase I sensitivity (see Supplementary Material of \citet{Degner_2012} for definition). For each sample, they then counted the number of DNase-seq reads mapping to each window,
standardized these counts by the total number of reads generated for each sample (to account for different read depths across individuals)
and used the resulting standardized counts as a molecular phenotype for association analyses.
 For each window in turn, they tested each nearby SNP for association with
 the DNase-seq data using a standard linear regression (after appropriate normalization, and controlling for confounding factors using 4 Principal Components).
  One analysis tested every SNP within 40,000 bases (40kb) of each window; another tested every SNP within 2kb.
  The first analysis identified 74,656 dsQTLs (FDR = 10\%) associated with 9,595 different windows. The second
  analysis identified 18,899 dsQTLs (FDR=10\%) associated with 7,088 different windows.

\subsection{Our analysis}

\citet{Degner_2012} observed  that typical dsQTLs affect chromatin accessibility over roughly 200-300bp. Based on this,
we decided to focus on slightly larger regions of size 1024bp for our wavelet-based association analyses (i.e., $B=1024$).
 From now on we refer to each 1024bp region as a 1024bp `site'. We focus our association analysis on the top 1\% of 1024bp sites with the highest DNase I sensitivity (in total 146,435 sites) selected as described in Appendix~\ref{Selection1024bp}. 
We focus on the top 1\% rather than the top 5\% as in \citet{Degner_2012} because \cite{Degner_2012} found that the majority of dsQTL are in the top 1\% of 100bp windows with the highest DNase I sensitivity. For each site, we use our wavelet-based hierarchical model, plus permutation,
described above, to obtain a $p$ value to test the null hypothesis, $H_0$: DNase-seq data at the site is unassociated with all near-by SNPs.
Here, we took ``near-by" to mean ``within 2kb of the site".

For comparison, we also implemented a testing approach analogous to the 100bp window-based approach from \cite{Degner_2012}.
In brief, we divided each 1024bp site into ten $\sim$100bp windows (nine of 100bp and one of 124bp). For each window
we computed a $p$ value for association of the DNase-seq data with each near-by SNP using standard linear regression as in \cite{Degner_2012}.
For this standard linear regression we quantile-normalized the phenotypes and corrected them for confounding factors using PCA, in the same way
as for the wavelet-based approach (Section \ref{normalization}). 
Then, we take the minimum of all these $p$ values (across all near-by SNPs and all 10 windows), $P_\text{min}$, as a test statistic of $H_0$.
We then assess the significance of $P_\text{min}$ by permutation, in the same way as we assess significance of our $\hat\Lambda_\text{max}$ by permutation (Section
\ref{permutation}). 



\subsection{A wavelet based approach increases power compared to a 100bp window approach}

To compare our wavelet-based approach with the window-based analyses, we applied both methods
to a subset of the data  (50,000 randomly selected 1024bp sites from the 146,435 sites). Each method yields a $p$ value
testing $H_0$ for each site. Using these $p$ values, we use the {\tt qvalue} package \citep{qvalue_package} to estimate the False Discovery Rate (FDR) for each method
at a given $p$ value threshold. We then compare the methods by the number of significant sites at a given FDR (more significant sites at a given
FDR being better).

\begin{figure}
{\small(a) } \\
\includegraphics[scale=0.5]{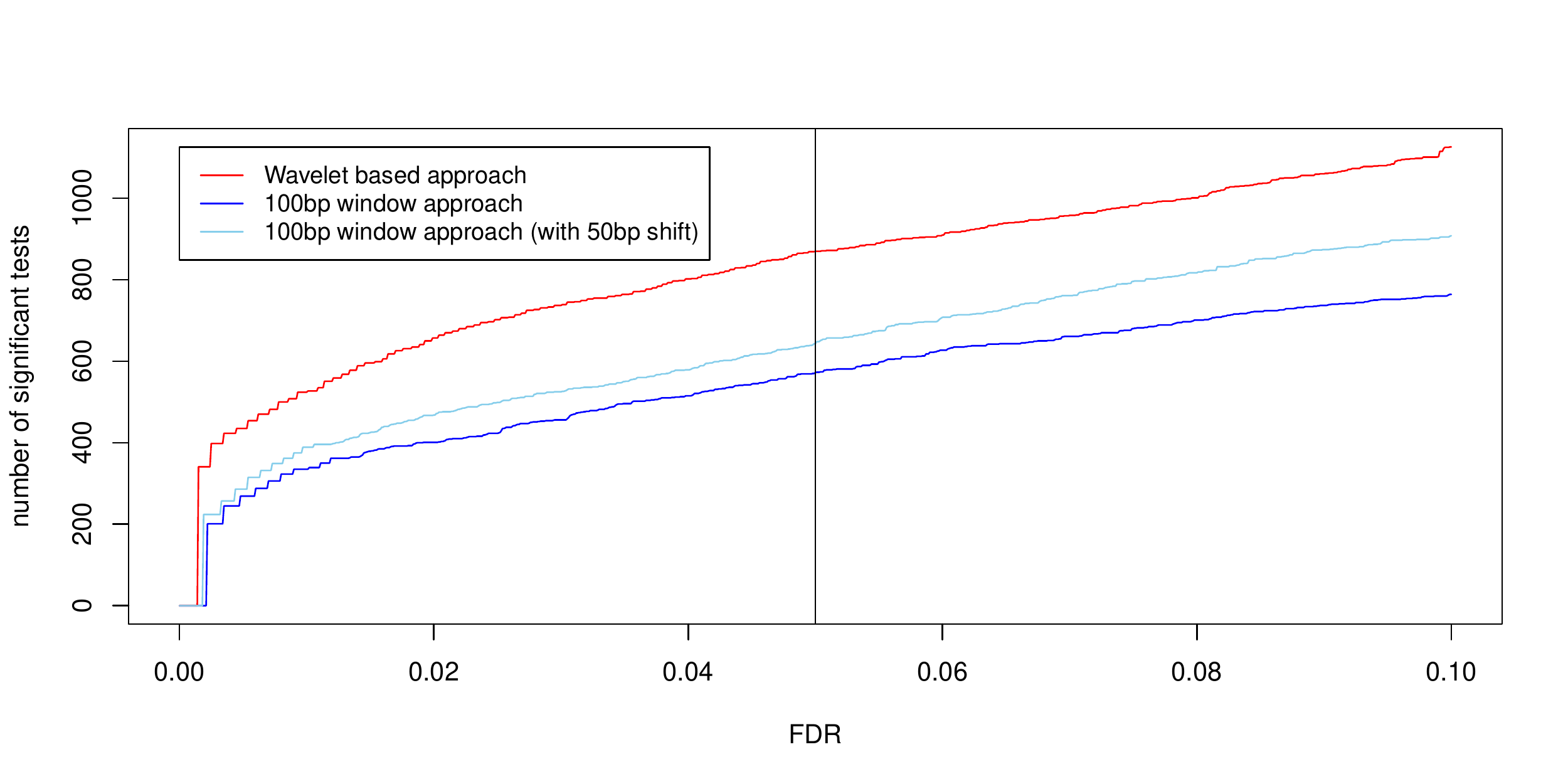}
{\small(b) } \\
\includegraphics[scale=0.5]{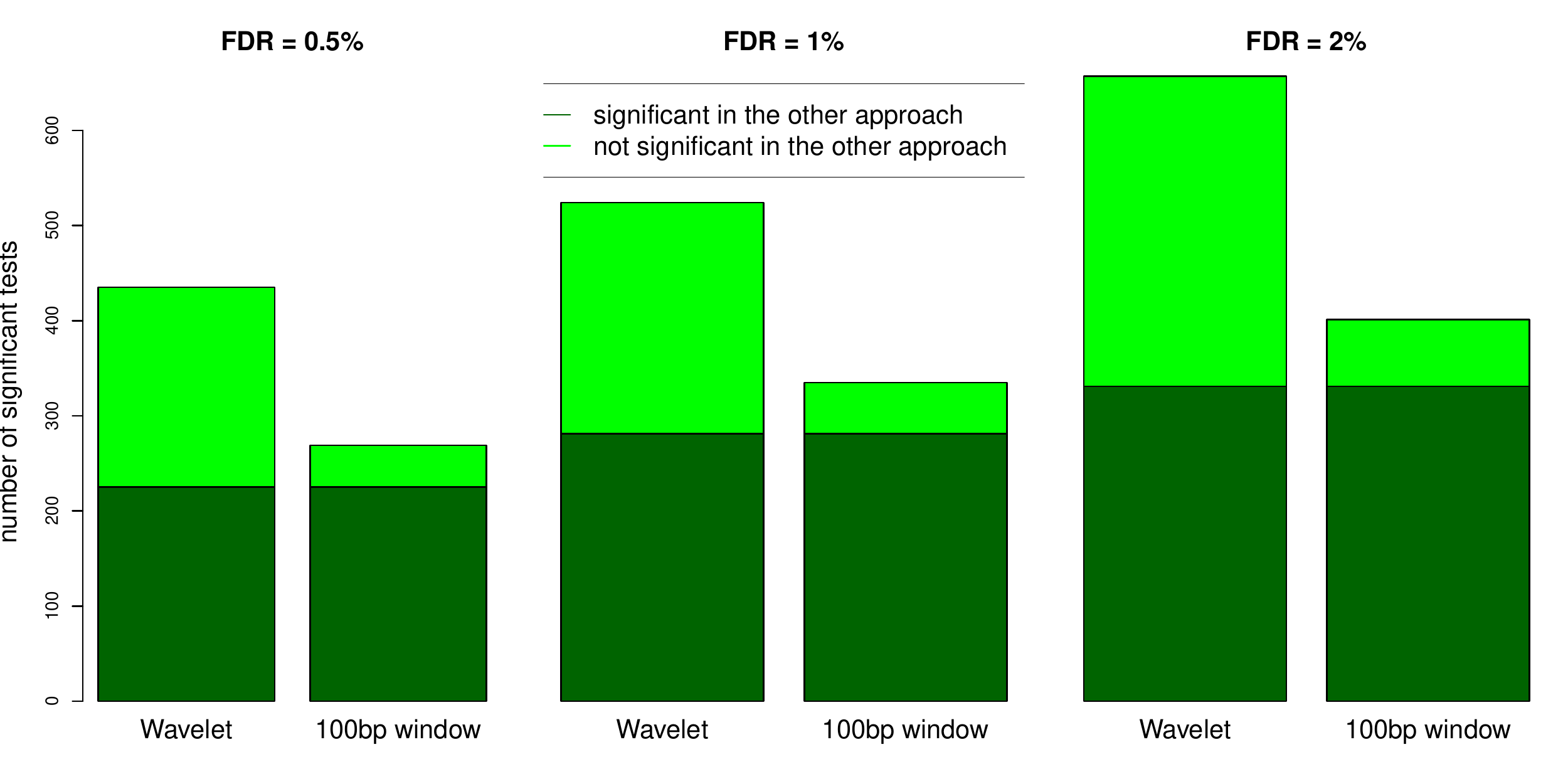}
\caption[]{\label{dsQTL_FDR:fig} {\bf The wavelet-based approach considerably increases power to identify dsQTLs compared to the 100bp window based approach.} 
Figure (a) shows the number of dsQTLs identified by each method at a given FDR. Black line indicates FDR of 0.05. Figure (b) shows the number of dsQTLs identified by the wavelet-based approach (Wavelet) and the 100bp window based approach (100bp window) at FDR of 0.005, 0.01, and 0.02. The number of dsQTLs identified by both approaches is highlighted by dark green.}
\end{figure}

Figure~\ref{dsQTL_FDR:fig} (a) compares the number of significant sites for each method as the FDR varies from 0.001 to 0.1. 
At all levels of the FDR the wavelet-based approach identifies considerably more significant sites than the 100bp window approach. 
For example, at FDR=0.05 the wavelet-based approach identifies 870 significant 
dsQTLs, compared with 572 dsQTLs for the 100bp window based approach, an increase of 52\%.
Moreover, most of dsQTLs detected by the 100bp window based analysis are also identified by the wavelet-based approach (Figure~\ref{dsQTL_FDR:fig} (b) , 84\%, 84\%, and 83\% for FDR of 0.005, 0.01, and 0.02, respectively).

To gain insights into commonalities and differences between the methods we manually examined effect size estimates for several examples.

Figure~\ref{dsQTL_wavelets:fig} (see also Supplementary Figure~\ref{supp1:fig} in Appendix~\ref{supp})  shows a typical example of a dsQTL identified
by both methods. These examples show a consistent strong effect across 200-300bp; consequently 
 at least one 100bp window fully overlaps the affected region, and the window analysis will successfully identify such examples, provided the effect is sufficiently strong. 
 
In contrast,  Figure \ref{dsQTL_only_wavelets:fig} shows two examples of dsQTLs identified by the wavelet analysis, but not the window-based analysis.
The dsQTL in Figure~\ref{dsQTL_only_wavelets:fig}(a) has a strong effect in a relatively narrow region (the strongest effect estimate in the second pink region spans $<$ 10bp).
The multi-scale nature of the wavelet approach makes it well adapted to detect this kind of narrow local feature, whereas the 100bp window analysis fails to capture it ($t$-statistic of the 100bp window containing the signal $\approx$2). This illustrates that the window based approach 
has limited power to identify signals that are very strong, but affect a region much smaller than the window size. 
 The dsQTL in Figure~\ref{dsQTL_only_wavelets:fig}(b) has a consistent effect spread over 200-300bp, qualitatively similar to typical dsQTLs identified by both methods.
However, the effect of this dsQTL is modest, and it fails to be significant in the window-based approach. Our explanation for this is that, being based
on 100bp windows, the window-based approach effectively uses only part (100bp) of the signal, whereas the 
multi-scale nature of the wavelet-based approach allows it to adapt to the scale of the signal, and make better use of the whole signal.
In summary, these examples illustrate how the window-based approach is inherently adapted to identifying effects that have a particular scale (100bp in this case)
and is suboptimal for effects that occur on either smaller scales (Figure \ref{dsQTL_only_wavelets:fig}a),  or larger scales (Figure \ref{dsQTL_only_wavelets:fig}b).

 Finally, Figure \ref{dsQTL_complex:fig} shows a slightly more complex example.
 This dsQTL shows  different effects in three regions: consistent in direction over about 100bp, in opposite directions over about 200bp, and modest in very narrow region. The 100bp window analysis misses the first signal, because no windows capture the whole signal, and misses the third one because it's modest and narrow. The third 100bp window fully overlaps with the second signal, but left and right sides of the window have effects in opposite directions and partially cancel each other out, resulting in a weak overall association. 

\begin{figure}
{\small chr17:10160989-10162012} \\
\includegraphics[scale=0.5]{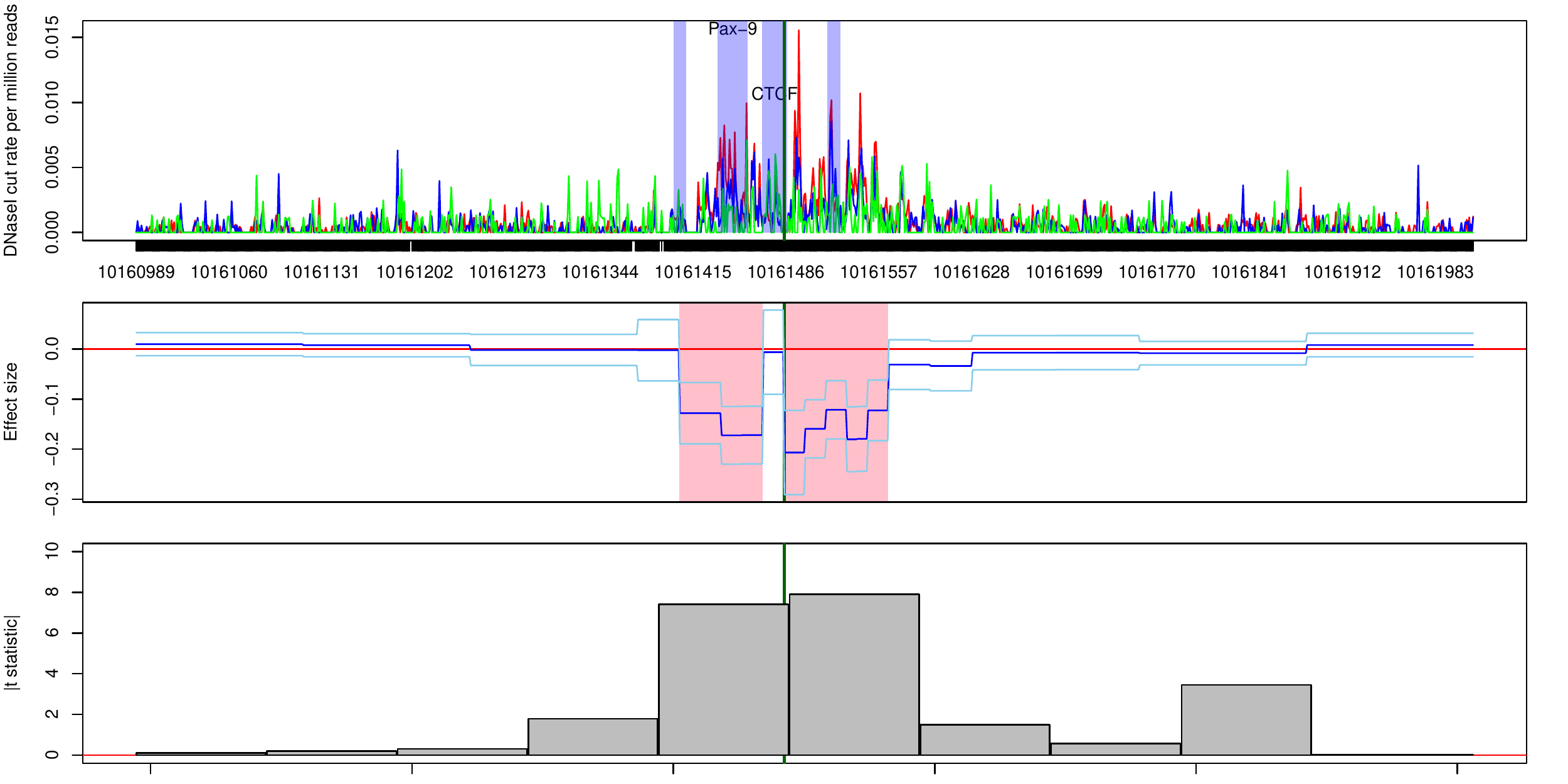}
\caption[]{\label{dsQTL_wavelets:fig} {\bf Example of typical dsQTL found by both methods.} 
The {\bf top panel} shows average DNase I cut rates along the site for each genotype class at the most strongly associated SNP (red=reference homozygotes; blue= heterozygotes; green=non-reference homozygotes). Dark green line indicates the position of the most strongly associated SNP. Purple blocks indicate putative transcription factor binding sites, identified using the software {\tt CENTIPEDE} \citep{Pique-regi_2011} (with a name on the top for known motifs). Black vertical lines below the x-axis indicate mappable bases (see Supplementary Material of \citet{Degner_2012} for definition). The {\bf middle panel} shows posterior mean for effect ($\alpha$) of this SNP  (blue), $\pm$3 posterior standard deviations (sky blue). Pink highlights regions showing strongest signal (zero is outside of mean $\pm$3 posterior standard deviations). The {\bf bottom panel} shows absolute value of t-statistic for each 100bp window.  The most strongly associated SNP: chr17.10161485 with minor allele frequency (MAF) of 0.39. For wavelet-based approach $\log\hat\Lambda_\text{max}= 73.09; p< 0.00001$. For window-based approach $p < 0.00001$.

}
\end{figure}

\begin{figure}
{\small(a) chr12:6264339-6265362} \\
\includegraphics[scale=0.5]{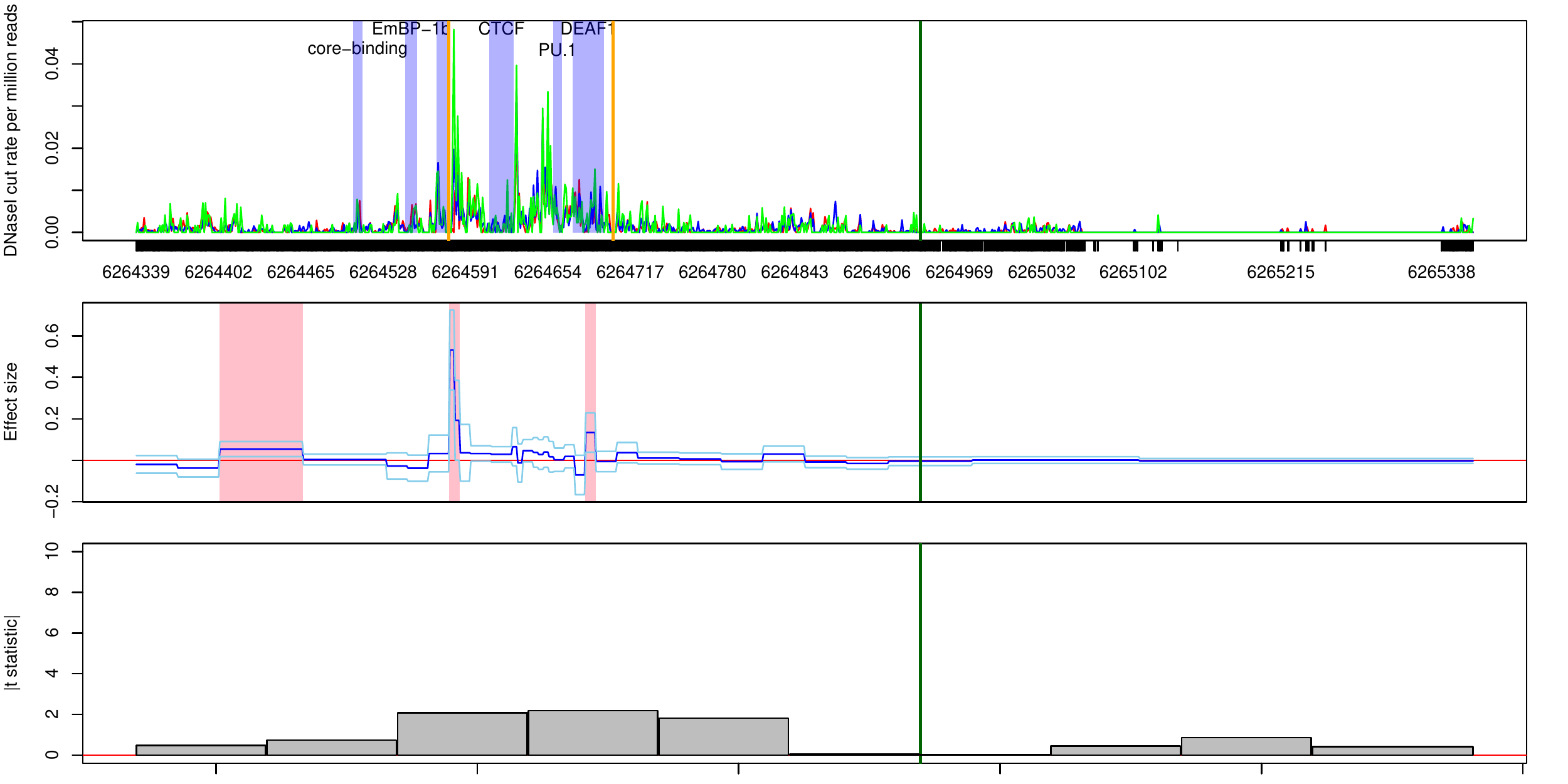}
{\small(b) chr10:59494639-59495662} \\
\includegraphics[scale=0.5]{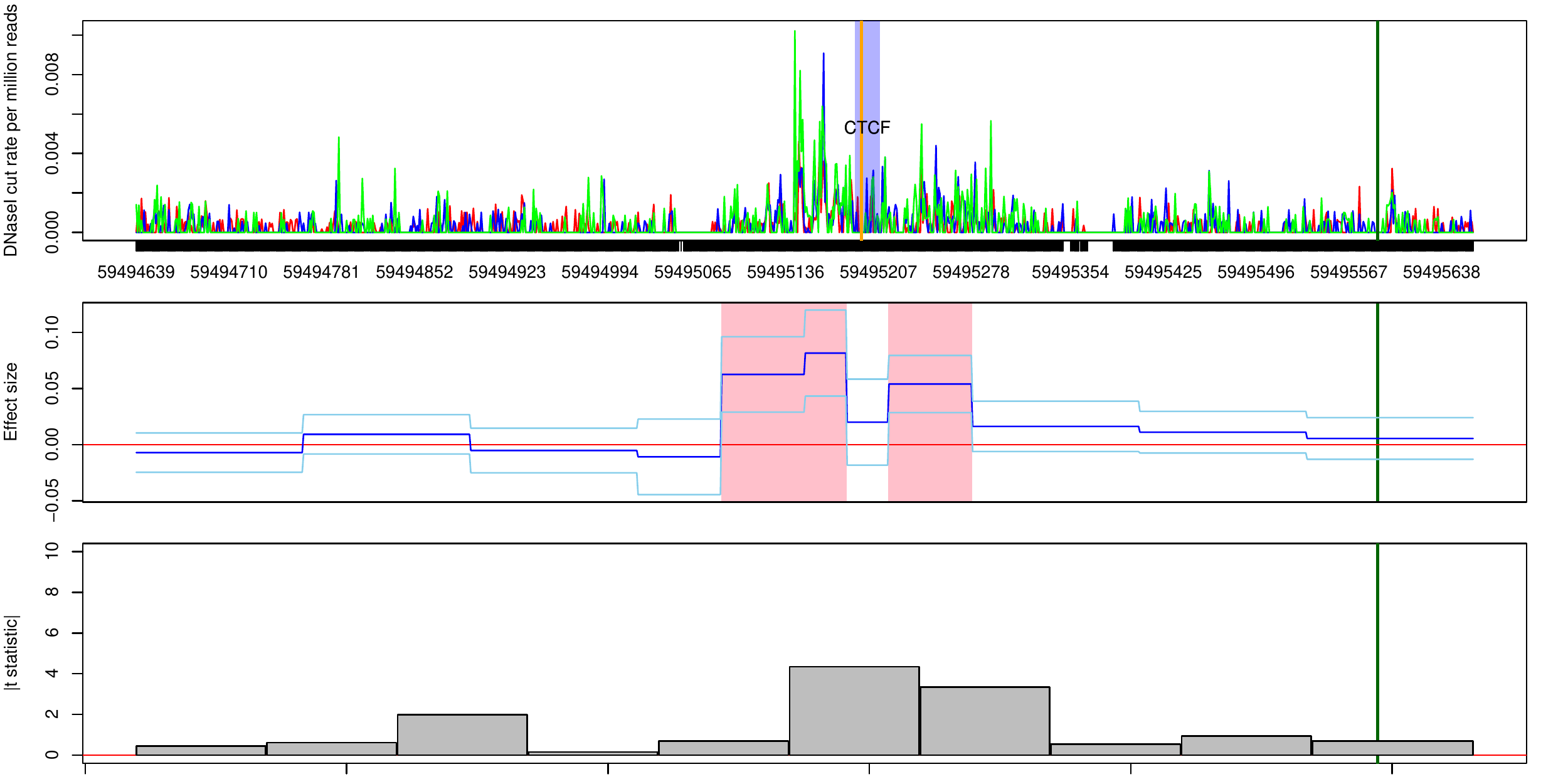}
\caption[]{\label{dsQTL_only_wavelets:fig} {\bf Examples of dsQTLs found by wavelet-based approach, but not by window-based approach.} 
Labels and colors are as in Figure~\ref{dsQTL_wavelets:fig}.
 (a) illustrates a dsQTL with a strong effect on a narrow region. The most strongly associated SNP: chr12.6264939 with MAF of 0.32. 
 For wavelet-based approach $\log\hat\Lambda_\text{max}= 25.97; p< 0.00001$. For window-based approach $p = 0.05$.
The two vertical orange lines indicate positions of two genetic variants that are in high linkage disequilibrium (i.e. highly correlated) with chr12.6264939.
(b) illustrates a dsQTL with modest effect over a larger region. The most strongly associated SNP: chr10.59495589 with MAF of 0.43. For wavelet-based approach
$\log \hat\Lambda_\text{max}= 14.11; p=0.0003$. For window-based approach $p = 0.01$. The orange line indicates the position of genetic variant that are in high linkage disequilibrium with chr10.59495589.}
\end{figure}

\begin{figure}
{\small chr2:110326889-110327912} \\
\includegraphics[scale=0.5]{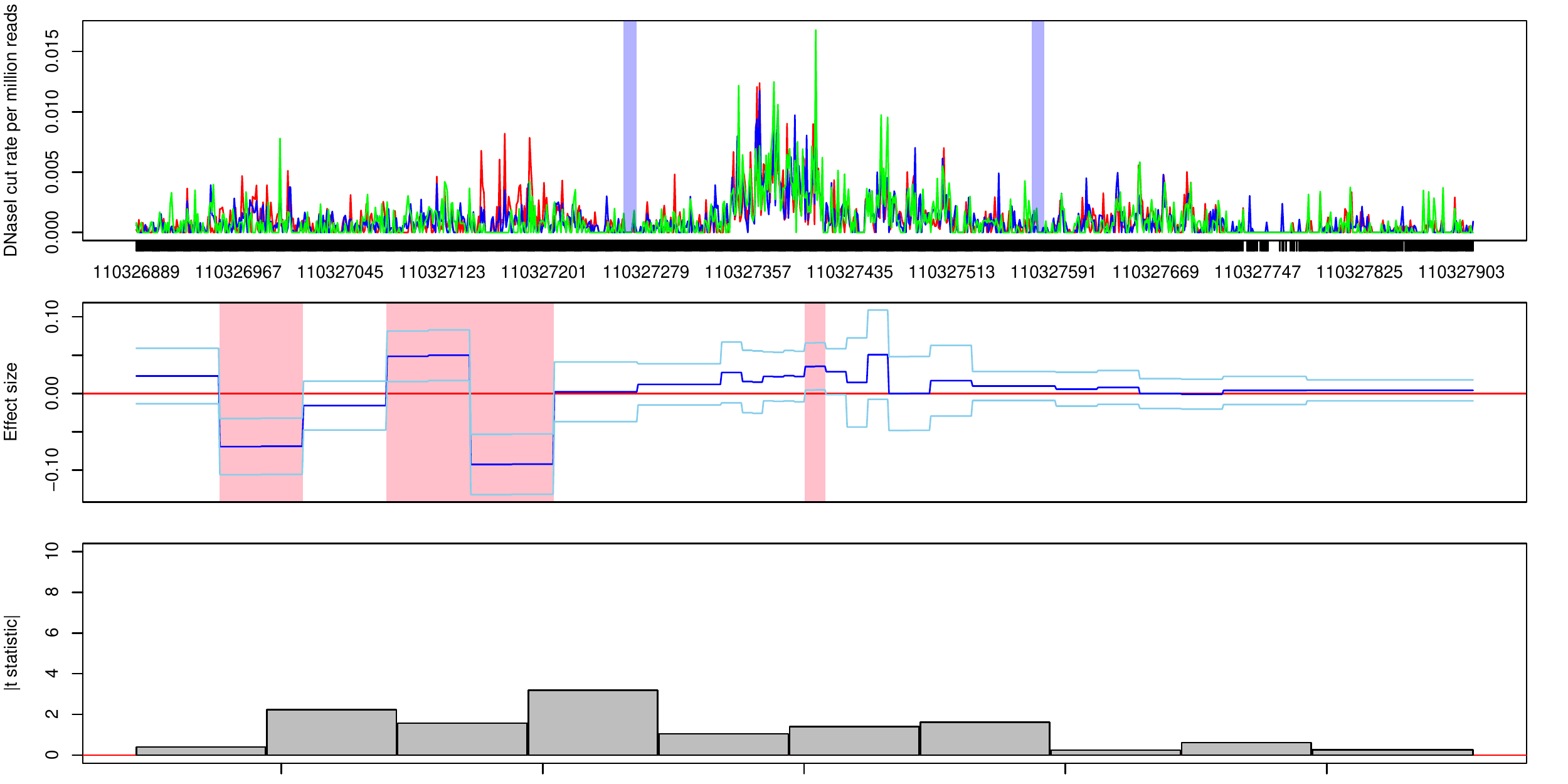}
\caption[]{\label{dsQTL_complex:fig} {\bf Example of dsQTL showing complex pattern of association with DNase I cut rates.} 
Labels and colors are as in Figure~\ref{dsQTL_wavelets:fig}. The most strongly associated SNP: chr2.110329846 with MAF of 0.43.
For wavelet-based approach
$\log \hat\Lambda_\text{max}= 22.01; p<0.00001$. For window-based approach $p = 0.23$. In this example the most strongly associated SNP is outside of the 1024bp site.}
\end{figure}

\subsubsection{Potential mechanism underlying dsQTLs}


It is possible that the different qualitative patterns of effect evident in the examples in Figures \ref{dsQTL_wavelets:fig}-\ref{dsQTL_complex:fig}
correspond to different functional mechanisms. 
With current data any discussion of mechanism is necessarily somewhat speculative. However, in some cases a putative mechanism is clearer
than others. In Figure~\ref{dsQTL_wavelets:fig}, the most strongly associated SNP  (green vertical line on Figure) is inside 
a binding site for CTCF (CCCTC binding factor), and the effect spans two regions either side of the binding site (each about 100bp highlighted by pink),
with the effect dropping to zero at the binding site itself. This effect exemplifies typical TF binding patterns, which often exhibit a distinct drop in 
DNase cut rates within TF binding sites \citep{Pique-regi_2011} (referred to as the DNase I ``footprint") because the binding of the TF ``protects" the DNA against
the cutting action of the DNase I enzyme. The effect estimate in Figure \ref{dsQTL_only_wavelets:fig} shows a similar footprint pattern around another CTCF binding site, and although the most strongly associated SNP is not in the CTCF binding site, another highly associated SNP is in that binding site (orange line; $r^2$ between these two SNPs is 0.9),
and this SNP seems more likely to be the actual functional variant. Thus, these two examples appear to share a common mechanism by which chromatin accessibility
is related to changes in CTCF binding. 

In contrast to these typical footprint patterns, the effect in Figure~\ref{dsQTL_only_wavelets:fig} (a) is quite different,
with one narrow region ($<$ 10bp) showing the biggest effect (the second pink region).
The most strongly associated SNP (green line) lies a few hundred base pairs from this strong effect, but
two other SNPs (orange vertical lines) that show almost identical association strength ($r^2>0.99$ with the strongest SNP) lie closer. One of these SNPs
lies in a putative TF binding site that coincides with
the narrow region of strongest effect. It seems plausible that this SNP is the functional variant influencing chromatin accessibility, and that the
changes in chromatin accessibility in this case are, as for the other examples, related to transcription factor binding. However, if so, the reason for
the effect being concentrated within the narrow area, rather than distributed around the TF binding site, is unclear.

Finally, the most strongly associated SNP in Figure~\ref{dsQTL_complex:fig} lies outside of the 1024bp window. The effect pattern here includes
almost-compensatory increases and decreases in chromatin accessibility, suggesting that the dsQTL is ``shifting" accessibility from some locations to others,
possibly by rearranging the underlying nucleosome positioning (although the precise mechanism by which this occurs is unclear).

\subsection{Shifting windows provides modest gain in power}

In some of the examples we examined (e.g.~Figure~\ref{dsQTL_complex:fig}), the 100bp window approach appeared to miss a signal because no single window fully overlapped the region affected by the dsQTL. This suggested that power might be increased by using overlapping, rather than non-overlapping, windows.
To assess this we modified the 100bp window approach to use 19 overlapping windows (the additional 9 windows being obtained
by shifting each of the first nine windows 50bp to the right). The test  statistic for this modified approach is the minimum $p$ value across 19 windows, 
and we assessed significance by permutation as before. We compared this modified 100bp window approach to the other two approaches by applying it to the 50,000 sites,
and computing the number of significant dsQTLs at a given FDR. As shown in Figure~\ref{dsQTL_FDR:fig} (a), it increases power compared with the non-overlapping windows, but remains well short of the wavelet-based approach. Looking at individual examples, we find the use of overlapping windows helps to identify the dsQTL in Figure~\ref{dsQTL_complex:fig} (p-value $<$ 0.00001) as the third 50bp-shifted window completely captures the signals that are consistent in direction over about 100bp (See Supplementary Figure~\ref{supp2:fig} in Appendix~\ref{supp}). However, it still missed both the dsQTLs in Figure~\ref{dsQTL_only_wavelets:fig}.

\subsection{A wavelet-based association analysis of the entire data set}

We next applied the wavelet-based approach to the full data set of 146,435 sites.
At an FDR of 10\% this yielded 3,176 sites with a dsQTL within 2kb.
Among these, 772 sites (24\%) are newly identified by the wavelet-based approach (i.e., not overlapping with the 7,088 100bp windows reported as having dsQTLs in 2kb {\it cis}-candidate region from \citet{Degner_2012}). 
We also found that 729 sites do not overlap with 1kb from the 7,088 100bp windows.

\subsubsection{Many dsQTLs affect expression levels of nearby genes}
A key finding of \citet{Degner_2012} was that the dsQTLs identified in their analysis were strongly enriched for being eQTLs: that is, being associated with changes in expression of at least one nearby gene. Specifically, using expression data on the same cell lines from \citet{Pickrell_2010}, they tested
their dsQTLs for association with expression. They found that
16\% of their dsQTLs are also significant  eQTLs (FDR = 10\%). These represent a very significant (450-fold) enrichment compared with random expectation. 
This is important because it suggests that altering chromatin accessibility and/or transcription factor binding may be a 
 common mechanism by which genetic variants influence gene expression. 

We therefore conducted a similar analysis for our dsQTLs, also using the data from \citet{Pickrell_2010}, and applying the methods from \citet{Degner_2012}
(see their Supplementary Material for details)
to the strongest associated SNP at each of the 3,176 significant sites identified in our analysis.
 We found that 19\% of dsQTL identified by the wavelet-based approach are also significant eQTLs (FDR = 10\%). 
 Among the 772 novel sites identified by the wavelet method, 15\% were also significant eQTLs. 
 The fact that these enrichments are similar to those reported in \citet{Degner_2012} suggests that the
 additional dsQTL sites we identified are likely to be reliable, rather than false positives. 

\section{Discussion}

 We have developed an effective and efficient statistical method for association analysis of functional data arising from high-throughput sequencing assays.
 This method, including permutation-based assessments of significance, is computationally tractable for genetic studies involving hundreds of thousands of tests.
 We applied our method to identify SNPs associated with chromatin accessibility, and demonstrated that the approach  can 
 exploit both high-resolution and lower-resolution features of the data to identify signals that are missed by
 simpler window-based approaches. Indeed, our analysis of data from \citet{Degner_2012} identified 772 novel putative dsQTLs not identified by the original analysis.
  
Although our methods were motivated primarily by genetic association studies for sequence-based molecular phenotypes, 
our approach is more general, and could also test for association between functional data and
other covariates, either continuous or discrete. For example, in a genomics context, it could be used to
 detect differences in gene expression (from RNA-seq data) or TF binding (from ChIP-seq data) measured on two groups (e.g. treatment conditions or cell types). 
 Or it could be used to associate a functional phenotype, such as chromatin accessibility, with a continuous covariate, such as ``overall" expression of a gene.
It could also be used for genome-wide association studies of functional phenotypes unrelated to sequencing.
 The main current limitation is that sample sizes should not be too small, since our Bayes Factor calculations, based on normal quantile-transformed
 data, will not work well for small samples. We have not experimented to determine adequate sample sizes, but in other settings we have
 found the quantile-transformed approach can work for sample sizes as small as 10 (M. Barber and M. Stephens, unpublished data). 
 We discuss modifying our approach to allow for smaller sample sizes below.
 
One of the most common assays now performed by sequencing is RNA-seq, and particular features of this assay merit special attention.
Specifically, because construction of mRNA effectively involves splicing together small parts of the gene (the ``exons"),
a proportion of the reads generated in an RNA-seq experiment will span splice junctions. These reads
naturally contain considerable information about splicing, but  this information is not captured in the information we use here (the first base
to which each read maps). Integrating the information in splice junction reads with our wavelet-based methods could be useful, but perhaps challenging.
On the other hand our method is not alone in failing to fully exploit splice reads,
and it also has some strengths that complement existing approaches to this problem. For example, it is common to 
use the number of reads mapping to ``known" exons as a phenotype to identifying SNPs that affect splicing \citep{Pickrell_2010}. This may work
well to identify certain types of effect (e.g. SNPs that affect whether or not an exon is spliced in), but less well for other effects
(e.g. extension of an exon beyond its usual boundaries). Because our method considers the shape of the read profile across the whole gene, without reference
to the ``known exons", it may be more effective at detecting this latter type of effect. 

To our knowledge, this is the first genetic association analysis that attempts to fully exploit high-resolution information from high-throughput sequencing assays.
As such there are many opportunities for potential improvements. First, our methods use a normal model for the (normal quantile-transformed) WCs,
and this transformation loses information. Particularly it loses the information that some WCs are based on small counts, and thus have
higher sampling variability than WCs based on larger counts. Here we partly addressed this issue by filtering out WCs based on low counts, but a more
principled approach may be expected to improve power. Further, as noted above, the normal quantile transformation requires moderate sample sizes. 
Both these issues could potentially be addressed by modeling the count nature of the sequence data directly, 
and we are currently experimenting with this approach, based on multiscale models for inhomogeneous Poisson processes \citep{Kolaczyk_1999, Timmermann_nowak_1999}.
Another possibility would be to consider transforms designed to allow wavelets to be applied to Poisson data \citep{Fryzlewicz2004}.
Second, we have here made use of Haar wavelets, and it may be that other wavelets will perform better. Indeed, the optimal choice of wavelets may be context-dependent. For example,
when applying wavelet denoising to ChIP-seq data on histone modifications \citet{Zhang2008} selected a wavelet known as Coiflet4, arguing that
its morphological characteristics are similar to the nucleosome peak shape. Our methods here could be directly applied with any choice of wavelet basis.

Finally, our hierarchical model assumes conditional independence of WCs (and effect sizes $\beta_{sl}$) given $\pi$ across scales and locations, and this conditional independence will not hold exactly in practice. Our approach partly addresses this issue by assessing significance of a test statistic by permutation, which gives valid p-values irrespective of whether modeling assumptions are correct. However, our procedure for estimating the shape of genotype effect still relies on the conditional independence assumption, and ultimately methods that exploit dependencies between the WCs should perform better. One way to model dependencies is to exploit the tree structure of WCs (and effect sizes $\beta_{sl}$) as described in \citet{Crouse1998}, and we are currently experimenting with this approach.

Software implementing the methods described here will be made available at \url{http://stephenslab.uchicago.edu/software.html}.

\section{Acknowledgments}
This work was supported by NIH grant HG02585.
We thank Jack Degner, Roger Pique-Regi and Jonathan Prichard for invaluable discussions and help with analyses of dsQTLs and Anil Raj, Ellen Leffler, and Sarah Urbut for helpful comments on an earlier version of the manuscript. We thank the members of the Pritchard, Przeworski,  Stephens, and Gilad labs for helpful discussions. 

\appendix

\section{Details of priors for Bayes Factor calculations}\label{DetailsBF}

We use the Bayes Factor calculations from \citet{Servin2007}, which are based on the following model and priors.
The model is:
\begin{equation}
y_{sl}^i = \mu_{sl} + \gamma_{sl} \beta_{sl} g^i + \epsilon_{sl}^i   \quad \text{with}  \quad \epsilon_{sl}^i \sim \mathcal{N}(0,  \sigma_{sl}^2).
\end{equation}
The priors are:
\begin{eqnarray}
\sigma_{sl}^2 &\sim& \Gamma^{-1}(\kappa_{sl}^a, \kappa_{sl}^b),\\
\mu_{sl}  | \sigma_{sl}^2 &\sim& N(0, \sigma_{\mu,sl}^2 \sigma_{sl}^2),\\
\beta_{sl} | \sigma_{sl}^2 &\sim& N(0, \sigma_{\beta,sl}^2 \sigma_{sl}^2)
\end{eqnarray}
and a discrete uniform prior on $\sigma_{\beta,sl} \in \{0.05,0.1,0.2,0.4\}$ (values which were chosen to span a wide range of potential effect sizes from very small to moderately large). As in \citet{Servin2007}, we use the limiting Bayes Factor obtained in the limits $\kappa_{sl}^a,\kappa_{sl}^b \rightarrow 0$, and $\sigma_{\mu,sl}^2 \rightarrow \infty$. In addition, when estimating effect sizes, we use the posterior distribution on $\beta_{sl}$ in the same limit. All posteriors we compute here are proper in these limits; see \citet{Servin2007} for discussion.

\section{Posterior distribution on effect sizes}\label{DetailsEffectSize}

\subsection{Posterior distribution of effect size on WC at scale $s$ and location $l$ when $\gamma_{sl}=0$, $\Pr(\beta_{sl} \given \gamma_{sl}=0, y_{sl}, g)$}\label{threeTdist}
In this section we drop the subscript $sl$ for convenience.
As derived in Protocol S1 in Supporting Information of \citet{Servin2007}, 
\begin{eqnarray}
	\sigma^2 \given y, g &\sim& \Gamma^{-1}( \frac{N+2\kappa^a}{2}, \frac{M}{2}),\\
	(\mu, \beta) \given y, g, \sigma^2 &\sim& N({\bf B},\sigma^2{\bf \Omega}),
\end{eqnarray}
where 
\begin{eqnarray}
	M &=& y^ty - {\bf B}^t{\bf \Omega}^{-1}{\bf B} + 2\kappa^b\\
	{\bf B} &=& {\bf \Omega} {\bf X}^t y \\ 
	{\bf \Omega} &=& ({\bf D}^{-1} + {\bf X}^t{\bf X})^{-1}\\
	{\bf D} &=& diag(\sigma_{\mu}^2, \sigma_{\beta}^2)
\end{eqnarray}
and {\bf X} has two columns, the first column being a vector of all ones and the second column being a vector of genotypes, $g$.
Then, 
\begin{eqnarray}
	\beta \given y, g, \sigma^2 &\sim& N({\bf B}_2,\sigma^2{\bf \Omega}_{22}),
\end{eqnarray}
where ${\bf B}_{2}$ denotes the second element of ${\bf B}$ and ${\bf \Omega}_{22}$ denotes (2,2)th element of ${\bf \Omega}$.

The posterior on effect size, $\Pr(\beta \given \gamma =0, y, g)$, can be written as
\begin{eqnarray}
	\int \int \Pr(\mu, \beta \given y, g, \sigma^2) \Pr(\sigma^2 \given y, g) \mathrm{d}\mu \mathrm{d}\sigma^2\\
	= \int \Pr(\beta \given y, g, \sigma^2) \Pr(\sigma^2 \given y, g) \mathrm{d}\sigma^2.
\end{eqnarray}
and 
\begin{eqnarray}
 &\propto& \int [\sigma^2]^{-\frac{1}{2}} \exp{[-\frac{(\beta-{\bf B}_2)^2}{2\sigma^2{\bf \Omega}_{22}}]  [\sigma^2]^{-\frac{N+2\kappa^a}{2}-1} \exp{[ -\frac{M}{2\sigma^2}   ]   }       }\mathrm{d}\sigma^2\\
 &\propto& [\frac{(\beta-{\bf B}_2)^2}{{\bf \Omega}_{22}} + M]^{-\frac{N+2\kappa^a+1}{2}}\\
 &\propto& [1 + \frac{(N+2\kappa^a)(\beta-{\bf B}_2)^2}{(N+2\kappa^a){\bf \Omega}_{22} M}    ]^{-\frac{N+2\kappa^a+1}{2}}.
\end{eqnarray}
Taking the limit $\kappa^a,\kappa^b \rightarrow 0$, and $\sigma_{\mu}^2 \rightarrow \infty$ yields the limiting posterior
\begin{eqnarray}
&\propto& [1 + \frac{N(\beta-{\bf B}_2^\ast)^2}{N{\bf \Omega}_{22}^\ast  (y^ty - ({\bf B}^{\ast})^t({\bf \Omega^\ast})^{-1}{\bf B}^\ast )}    ]^{-\frac{N+1}{2}},
\end{eqnarray}
where ${\bf B}^\ast$ and ${\bf \Omega}^\ast$ are obtained in the limit.
Then, the limiting posterior, $\Pr(\beta \given \gamma =0, y, g)$ is a three parameter version of a $t$ distribution \citep{Jackman2009} with density 
\begin{eqnarray}
	p(x | \nu,a,b) = \frac{\Gamma(\frac{\nu + 1}{2})}{\Gamma(\frac{\nu}{2})\sqrt{\pi\nu b}} \left(1+\frac{1}{\nu}\frac{(x-a)^2}{b}\right)^{-\frac{\nu+1}{2}}
\end{eqnarray}
where 
\begin{eqnarray}
	a &=& {\bf B}_2^\ast\\
	b &=& \frac{{\bf \Omega}_{22}^\ast  (y^ty - ({\bf B}^{\ast})^t({\bf \Omega^\ast})^{-1}{\bf B}^\ast )  }{N}\\
	\nu &=& N
\end{eqnarray}
and
\begin{eqnarray}
\mathrm{E}(x) &=& a \quad \text{for} \quad \nu > 1\\
\mathrm{Var}(x) &=& \frac{b\nu}{\nu-2} \quad \text{for} \quad \nu > 2.
\end{eqnarray}

\subsection{Posterior distribution of effect size on WC at scale $s$ and location $l$, $\Pr(\beta_{sl} \given y_{sl}, g)$}
The posterior distribution of the effect size on WC at scale $s$ and location $l$, $\Pr(\beta_{sl} \given y_{sl}, g)$ , can be written as
\begin{eqnarray}
	\Pr(\beta_{sl} \given \gamma_{sl}=1, y_{sl}, g)  \phi_{sl} + \Pr(\beta_{sl} \given \gamma_{sl}=0, y_{sl}, g)  (1-\phi_{sl}),
\end{eqnarray}
where $\Pr(\beta_{sl} \given \gamma_{sl}=0, y_{sl}, g)$ is a point mass at zero and $\phi_{sl}$ is 
\begin{eqnarray}
	&& \Pr(\gamma_{sl} = 1 \given \hat\pi , y_{sl}, g)\\
               &=&\frac{\Pr(y_{sl} \given g, \gamma_{sl} = 1)\Pr(\gamma_{sl} = 1 \given \hat\pi)}{   \Pr(y_{sl} \given g, \gamma_{sl} = 1)\Pr(\gamma_{sl} = 1 \given \hat\pi) + \Pr(y_{sl} \given g, \gamma_{sl} = 0)\Pr(\gamma_{sl} = 0 \given \hat\pi)}\\
               &=&\frac{\hat\pi_{s} \BF_{sl}}{\hat\pi_{s} \BF_{sl} + (1 - \hat\pi_{s})}.
\end{eqnarray}
Therefore, the posterior distribution of the effect size on WC at scale $s$ and location $l$ is a mixture of a point mass at zero and a three parameter version of a $t$ distribution with a location parameter $a_{sl}$, a squared of scale parameter $b_{sl}$, and the number of degrees of freedom $\nu_{sl}$  (see Appendix~\ref{threeTdist}). The three parameter $t$ distribution mixture proportion is $\phi_{sl}$.

\subsection{Effect size in the original space}
From $\alpha=W^{-1}\beta$, and noting that $W$ is orthogonal so $W^{-1} = W'$, the posterior mean and variance of $\alpha$ are given by
\begin{eqnarray}
\mathrm{E}(\alpha_b) &=& \displaystyle\sum_{s, l} a_{sl}\phi_{sl}w_{sl,b},\\
\mathrm{Var}(\alpha_b) &=& \displaystyle\sum_{s, l} [v_{sl} + a_{sl}^2 - a_{sl}^2\phi_{sl} ] \phi_{sl}w_{sl,b}^2,\\
      v_{sl}  &=& \frac{b_{sl}\nu_{sl}}{\nu_{sl} -2}.
\end{eqnarray}
where $w_{sl,b}$ is the element of the DWT matrix in row corresponding to scale $s$ and location $l$, and the column corresponding to base $b$.

\section{Selection of the top 1\% of 1024bp sites with the highest DNase I sensitivity}\label{Selection1024bp}
We focus our association analysis on the top 1\% of 1024bp sites with the highest DNase I sensitivity (146,435 sites) that are selected by the following procedure.  
As in \citet{Degner_2012}, we divide the whole genome into non-overlapping 100bp windows and rank them according to a DNase I sensitivity (for the definition of DNase I sensitivity, see Supplementary Material of \citet{Degner_2012}). 
Then, we select the top 1\% of 100bp windows with the highest DNase I sensitivity. 
We merge those 100bp windows if they are adjacent to each other and if the length of the merged window is less than 1024bp, leading to 146,435 windows. The 146,435 1024bp sites used in our analysis are centered at those windows.

\section{Supplementary Figures}\label{supp}

\begin{figure}
{\small chr8:8463639-8464662} \\
\includegraphics[scale=0.5]{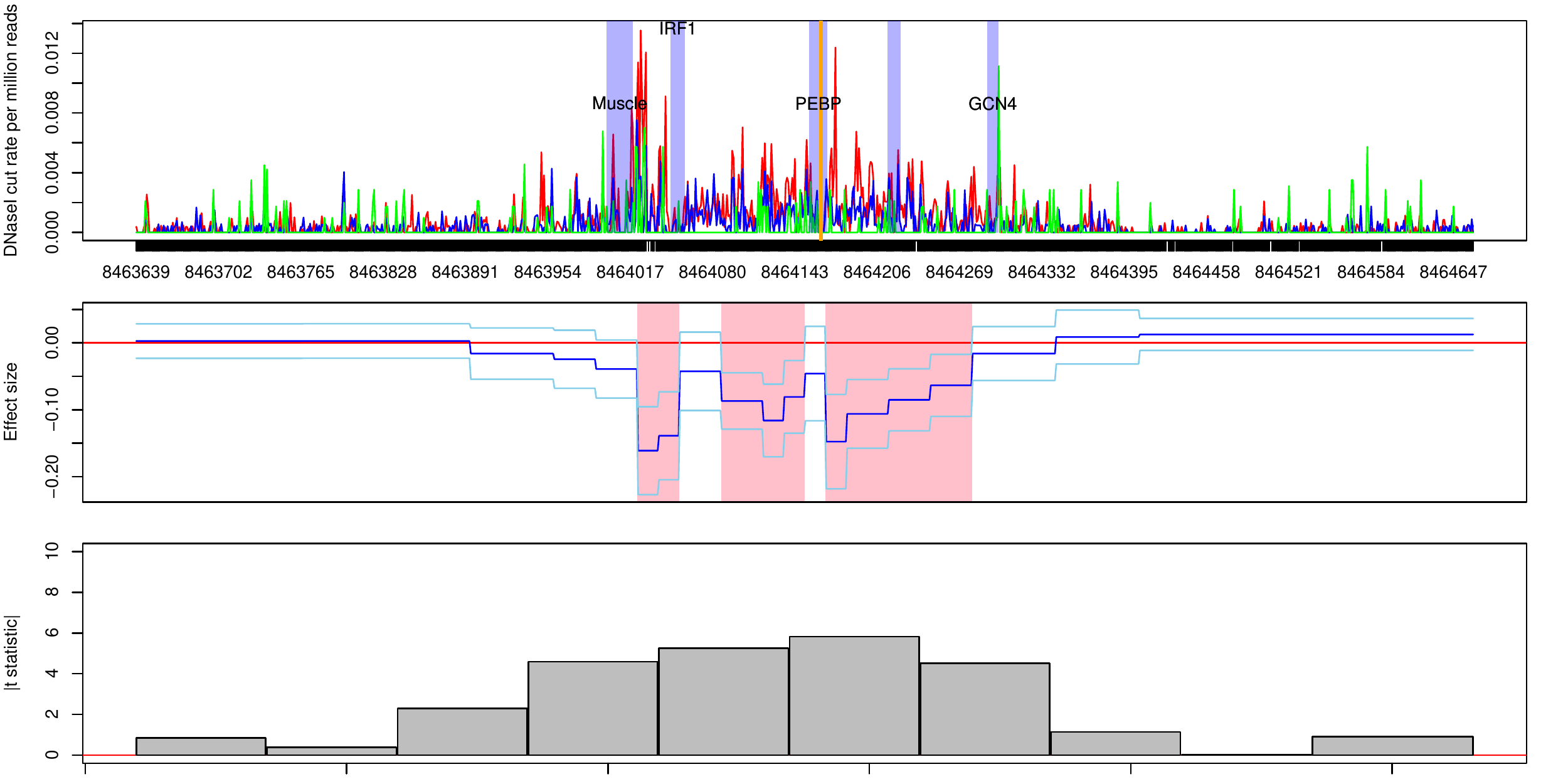}
\caption[]{\label{supp1:fig} {\bf Example of typical dsQTL found by both methods.} 
Labels and colors are as in Figure 2 of the main text. 
The most strongly associated SNP: chr8.8462948 with MAF of 0.29. For wavelet-based approach $\log\hat\Lambda_\text{max}= 39.89; p< 0.00001$. For window-based approach $p < 0.0001$.
The orange line indicates the position of genetic variant that are in high linkage disequilibrium with chr8.8462948 ($r^2 >$ 0.99).
}
\end{figure}

\begin{figure}
{\small chr2:110326889-110327912} \\
\includegraphics[scale=0.5]{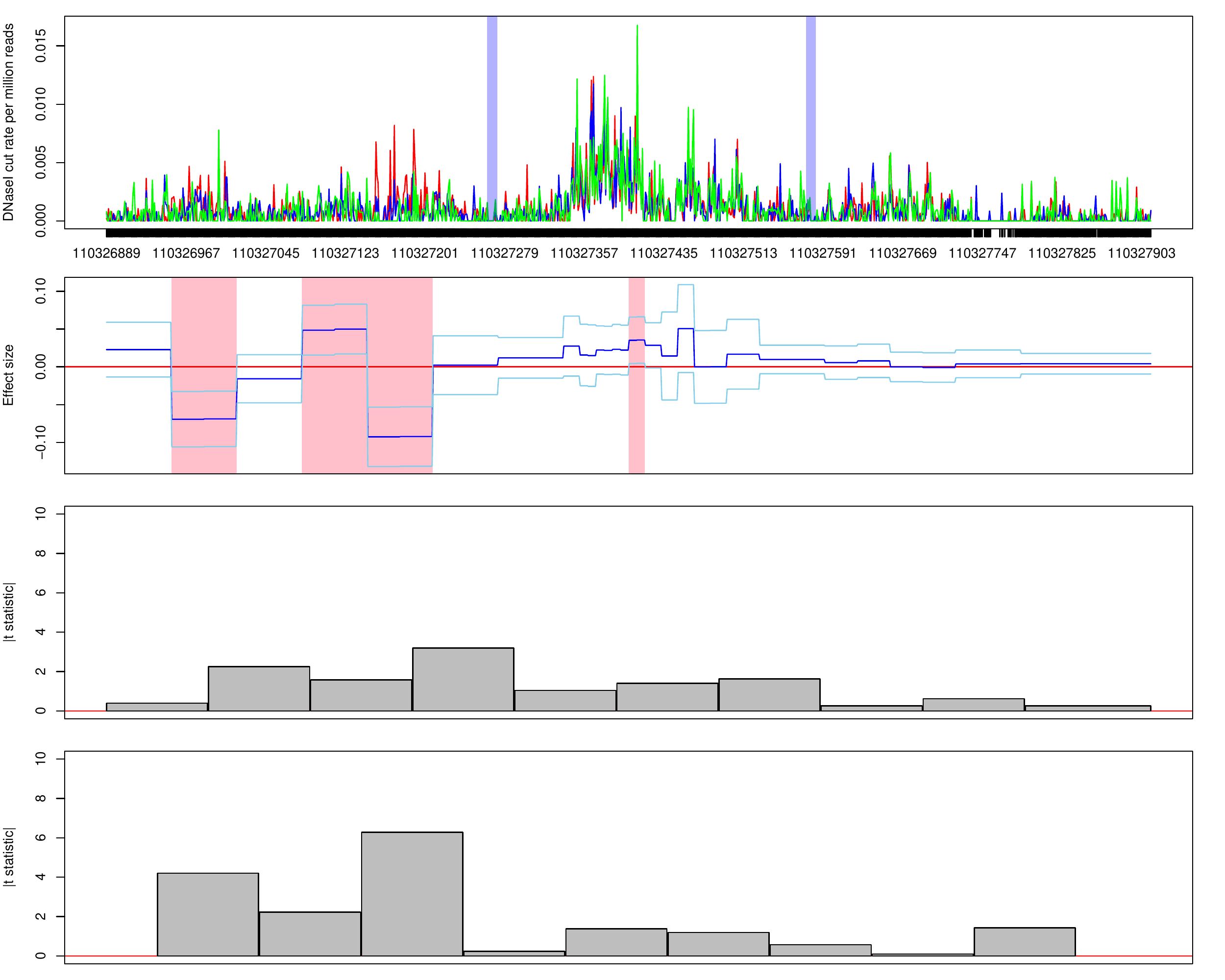}
\caption[]{\label{supp2:fig} {\bf Example of dsQTL showing complex pattern of association with DNase I cut rates.} 
Labels and colors are as in Figure 2 of the main text. 
The bottom figure shows absolute value of t-statistic for each 50bp-shifted 100bp window. 
p $<$ 0.00001 (wavelet-based approach and 100bp window approach with 50bp shift)  0.23 (100bp window approach) } 
\end{figure}


\bibliographystyle{imsart-nameyear} 
\bibliography{wavelets}

\end{document}